# Model Gateway: Management Platform for Model-Driven Drug Discovery

Yan-Shiun Wu[a,*], Sai Mahit Vaddadi[a], Zachary A. Rollins[a], Nathan A. Morin[a,*]

[a]*Eli Lilly and Company, Lilly Corporate Center, Indianapolis, IN 46285, USA*

**Abstract**

Pharmaceutical drug discovery demands machine learning (ML) infrastructure that goes beyond general-purpose Machine Learning Operations (MLOps): inference-time composition of multiple models for multi-parameter optimization (MPO), version management for physics-based models without serialized ML artifacts, enterprise compound library precomputation, and governance structured around scientific organizational units rather than generic access controls. No existing commercial or open-source platform simultaneously addresses this full set of requirements. This paper presents the Model Gateway, a cloud-based platform for managing machine learning and scientific computational models across drug discovery pipelines, providing centralized version control, pharma-structured governance, asynchronous execution, consensus model orchestration, automated retraining, and a unified application programming interface (API) service for heterogeneous clients including molecular design suites and Large Language Model (LLM) agents. In production at Eli Lilly, the platform governs more than 200 deployed models spanning small molecule, peptide, and antibody modalities and serves more than five downstream applications across all phases of the Design–Make–Test–Analyze cycle.

## Introduction

Machine Learning Operations (MLOps) has emerged as a critical discipline for operationalizing machine learning (ML) models in production environments, yet remains a nascent and fast-evolving field [1,2]. MLOps extends traditional DevOps practices by incorporating the unique challenges of managing code, data, and models [3], introducing Continuous Training alongside Continuous Integration/Continuous Delivery (CI/CD) to address model drift [4]. Despite growing adoption, fully effective MLOps frameworks remain scarce [2], and organizations face significant technical debt deploying models without a platform to systematically manage them at scale [5].

*Corresponding author

*Email addresses:* Allen.Wu@lilly.com (Yan-Shiun Wu), morin_nathan_a@lilly.com (Nathan A. Morin)

**From DevOps to MLOps: a brief primer**

*DevOps* unifies software development and IT operations through automated build, test, and deployment pipelines (CI/CD), version control, and infrastructure as code, so that application changes can be released frequently and reliably. *MLOps* extends this discipline to the data- and model-centric realities of machine learning: in addition to versioning code, teams must version *datasets*, *trained model artifacts*, and *evaluation results*, and pipelines must support *Continuous Training*—periodically retraining models as new data becomes available—and *model monitoring* to detect accuracy drift in production. In drug discovery, this means a successful platform must reproducibly track not only source code, but also the experimental assay snapshot a model was trained on, the resulting weights, the governance metadata describing who can call it, and the audit log of every inference request.

In pharmaceutical drug discovery, these challenges are compounded by domain-specific complexities. Drug discovery operates through iterative Design-Make-Test-Analyze (DMTA) cycles [5,6], where ML models serve dual roles: project-specific models trained on evolving experimental data and global models for properties such as toxicity and off-target effects [5]. Only a small percentage of ML projects in drug discovery reach production [5], and the industry is only beginning to define MLOps best practices [6]. Models must increasingly span diverse chemical matter across the industry—small molecules, peptides, antibodies, RNA, DNA, and complex conjugates [6]—and as organizations demand faster DMTA cycles with "predict first" paradigms, model management has become a critical operational bottleneck[6].

The Model Gateway platform addresses these challenges by providing a cloud-based, production-grade infrastructure for end-to-end model lifecycle management in drug discovery. Built on cloud-native technologies and MLOps best practices, the platform enables centralized model deployment, versioning, monitoring, and integration across diverse clients and discovery workflows, aiming to reduce technical debt, increase model accessibility, ensure reproducibility, and accelerate the translation of ML innovation into drug discovery impact. The Model Gateway is a proprietary internal platform developed and operated within Eli Lilly and Company; the architecture, design rationale, and operational learnings are described here as a reference experience report rather than a release of source code, in the spirit of recent published accounts of pharma-internal ML platforms[6] and emerging open-source analogs[7].

### Build versus Buy: Motivating Design Choices

Three categories of existing solutions were evaluated and found insufficient in different ways. MLOps platforms (Amazon SageMaker, Databricks, Azure ML, Domino) provide no mechanism for inference-time consensus and multi-parameter optimization (MPO) composition across models, no registry concept for physics-based models without serialized ML artifacts, and no provision for pre-scoring enterprise compound libraries as a first-class operational pattern. Vendor scientific suites (Schrödinger LiveDesign, Dotmat-ics Luma, BIOVIA Pipeline Pilot) are designed as scientist-facing frontends rather than neutral backend APIs; meaning any one the model-serving layer would privilege a single vendor and prevent competing design environments from accessing the same catalog through a common interface. Open-source analogs (NERDD[7], QSARtuna[8], ADMET-AI[9]) address individual capabilities—microservice deployment, automated quantitative structure-activity relationship (QSAR) modeling, and pretrained absorption, distribution, metabolism, excretion, and toxicity (ADMET) prediction—but none simultaneously provides pharma-structured role-based access control (RBAC), asynchronous execution, consensus orchestration, automated retraining against assay snapshots, and agentic LLM access; Table 1 shows this gap across the full landscape.

Model Gateway was therefore built to close that gap, composing mature open-source infrastructure (Kubernetes, MLflow, Redis, Nextflow) where sufficient and building only what did not exist. Internal ownership also preserves access to platform internals—queue state, execution history, per-model usage telemetry—needed for future extensions that vendor API surfaces cannot expose: pharma-specific monitoring layers and predictive container warming based on learned DMTA usage patterns. Because no single vendor covers the full pharma MLOps stack, assembling equivalent coverage from partial solutions would introduce integration seams at every boundary with simultaneous lock-in across all of them. Internal ownership is particularly consequential given the pace of change in adjacent verticals: LLM agent frameworks and AI-native molecular design suites are both evolving fast enough that no vendor integration surface is yet stable. Model Gateway can adapt its tool surface, metadata schema, and API contracts as those ecosystems mature, ensuring Lilly can adapt quickly to novel model capabilities while exposing model insights to the full range of vendor and external scientific tools used across the enterprise.

### Related Systems and Competitive Landscape

Published descriptions of pharma-internal MLOps platforms remain rare; aside from AstraZeneca's DMTA platform [6], most peer organizations have disclosed their internal infrastructure only obliquely [10] or through industry consortia such as Polaris [11]. Domain-specific computational workflows such as Admiral [12]—an automated molecular dynamics and analysis pipeline—represent a complementary class of pharma-internal tooling focused on orchestrating physics-based simulations rather than governing a heterogeneous model catalog; such tools are consumers of a governed serving backend rather than alternatives to one. Open-source analogs have begun to emerge in the academic cheminformatics community—notably the NERDD microservices framework deployed at the University of Vienna [7]—and adjacent open-source chemistry-AI tooling (REINVENT 4[13], QSARtuna [8], ADMET-AI [9]) provide individual capabilities that Model Gateway integrates rather than replaces. In parallel, a rapid expansion of agentic chemistry systems (ChemCrow [14], Coscientist [15], CACTUS [16], Ock et al. [17], Auditable Agent Platform [18], AstraZeneca's modularity study[19]) has demonstrated that LLM agents are now a credible interface to predictive infrastructure;

these systems are consumers of a governed model-serving backend rather than alternatives to one, and none publishes an account of an *enterprise* platform that simultaneously serves them with governed, auditable, asynchronous model execution. Commercially, AI-native drug discovery platforms such as Chemistry42 [20] (Insilico Medicine) represent a growing class of vertically-integrated frontends that bundle generative design, ADMET prediction, and molecular dynamics within a single licensed application; they are not designed to function as vendor-neutral model-serving backends callable across heterogeneous discovery environments. NVIDIA BioNeMo provides GPU-optimized biomolecular model training and customization infrastructure alongside NVIDIA Inference Microservices (NIM) for self-hosted inference, and supports proprietary model training on user data; however, it is positioned as a developer framework rather than a governed enterprise platform and lacks documented support for organizational RBAC or automated retraining pipelines triggered by new experimental data. Table 1 summarizes the resulting landscape and positions Model Gateway against representative published and commercial systems.

## Methodology

The Model Gateway (Fig. 1) platform comprises three layers: client applications (including LLM agents and molecular design suites), the Model Gateway infrastructure itself, and the underlying computational models. Built on cloud-native technologies, the platform leverages Kubernetes [27] for container orchestration, Docker [28] for containerization, and MLflow [29] for model registry and artifact management. The platform architecture is organized into five major pillars:

A. Model Owner Control Panel

B. Platform Administration Tools

C. Model Execution

D. Automated Model Retraining

E. Client Integrations

Table 2 summarizes the component technology selections underlying these pillars, documenting the options evaluated and the rationale for each choice as a reference for peer organizations.

### *Model Owner Control Panel*

The Model Owner Control Panel provides a web-based interface enabling computational scientists and domain experts to manage their models through three core capabilities: hierarchical access control, flexible version management with MLflow integration, and metadata management for large language model (LLM) agent discovery and client routing.

*Access Control*

The model access control feature provides role-based access control (RBAC) across all facets of the platform, integrated into enterprise and agentic identity sources, enabling model governance across heterogeneous discovery organizations. The platform implements a three-tier access control scheme:

- **Superuser**: Platform administrators with full access to all models and administrative functions, responsible for initial model provisioning and cross-organizational governance
- **Model Owner**: Domain experts and computational scientists with full control over their models including versioning, metadata management, and access delegation
- **Model Reader**: End users and client applications granted read-only access to specific models

Models are private by default, requiring explicit grants. Model Owners manage Reader access based on organizational structure—mapping models to discovery programs, therapeutic areas, or computational groups—ensuring sensitive models remain isolated to authorized organizations while broadly applicable models can be shared across the enterprise.

*Version Management*

The Model Gateway implements a flexible version management architecture leveraging MLflow [29] as the primary source for tracking ML model versions, providing capabilities for managing model stages (staging, production, archived) and maintaining artifacts across the ML lifecycle. Recognizing that pharmaceutical discovery encompasses models beyond traditional ML—including physics-based simulations and cheminformatics algorithms—the platform also maintains a custom version management system for non-standard models, ensuring comprehensive version control across heterogeneous model types.

*Metadata Management*

The model metadata system serves as the central information registry for computational models, providing structured data for automated model discovery, intelligent routing, and access enforcement. The metadata schema includes model identifiers, strongly typed input/output schemas, natural language descriptions, version information, backend resource requirements, and access controls. Model descriptions are critical for LLM agent integration, enabling agents to semantically match user requests to model capabilities, while also supporting human-readable interfaces for client applications.

*Platform Administration Tools*

Platform Administration Tools provide operational visibility through interactive API documentation (Swagger [36]), audit logging of all inference requests to S3 for regulatory compliance, and real-time job queue monitoring for performance analysis. These capabilities support both developer integration and ongoing platform governance.

*Model Execution*

Model Execution provides the runtime infrastructure for executing computational models at scale through asynchronous job processing, elastic scaling on Amazon Elastic Kubernetes Service (EKS), and flexible consensus model aggregation.

*Asynchronous Request and Retrieval*

Asynchronous model execution addresses variable computational latency in drug discovery, where execution times range from milliseconds for property predictions to hours for molecular dynamics simulations. The platform implements a job queue architecture where clients submit requests and receive immediate workflow UUIDs, decoupling submission from retrieval. Upon submission, requests are persisted to a Redis Cluster job queue with a custom prioritization algorithm balancing user priority and submission history. Worker processes execute models in isolated containers and store results in Redis with UUID-indexed retrieval, enabling clients to poll for completed results while long-running computations execute in the background.

*Elastic Model Scaling*

Elastic model scaling enables cost-effective adaptation to fluctuating workloads through horizontal and vertical strategies. The platform leverages Amazon EKS [31] for container orchestration and implements Kubernetes Event-driven Autoscaling (KEDA) [32], which monitors Redis queue depth to dynamically provision or deprovision worker replicas. Vertical scaling complements this by adjusting resources allocated to individual containers—supporting heterogeneous graphics processing unit (GPU) and central processing unit (CPU) instance types—ensuring computationally intensive models receive adequate capacity without over-provisioning for lightweight models.

*Dynamic Consensus Model Management*

Dynamic consensus model management enables computational scientists to create composite models that aggregate predictions from multiple individual models, improving accuracy through ensemble methods. Drug discovery frequently requires balancing competing molecular properties—potency, selectivity, solubility, toxicity—where consensus modeling provides systematic compound prioritization. The platform supports weighted arithmetic combinations and normalization methodologies from multiparameter optimization (MPO) frameworks [37], enabling Model Owners to define aggregation formulas across child models as illustrated in Alg. 1. Parallel orchestration of constituent models at inference time enables rapid prototyping and deployment of ensemble strategies without custom backends.

$$\begin{aligned} &Result = ChildModel1 + ChildModel2, && \text{if } ChildModel1 \leq 0 \\ &Result = ChildModel1 \times ChildModel2, && \text{otherwise.} \end{aligned}$$

**Algorithm 1** Simple Consensus Model Configuration

```
"ConsensusModel":
  childModels:
    ChildModel1:
      backend: ChildModel1_Backend
      name: "ChildModel1_Name"
    ChildModel2:
      backend: ChildModel2_Backend
      name: "ChildModel2_Name"
  formula: >
    "ChildModel1 <= 0 ?
      ChildModel1+ChildModel2 :
      ChildModel1*ChildModel2"
```

*Automated Model Retraining*

Automated model retraining provides continuous model improvement by orchestrating end-to-end ML pipelines that automatically incorporate newly generated experimental data into updated model versions. As illustrated in Fig. 2, the Model Training component processes research data through training workflows, containerizes trained models in the Model Foundry, and deploys them via Helm to the Kubernetes cluster, where they are discoverable through the Model Gateway Core.

The platform leverages Amazon Web Services (AWS) cloud infrastructure to implement scalable retraining workflows triggered on configurable schedules (e.g., monthly, quarterly), pulling molecular data from corporate databases, executing training pipelines, and deploying updated models. The retraining infrastructure utilizes Nextflow [34] to automate data extraction, model training, validation against test sets, and deployment upon meeting performance criteria. When a new version is successfully produced, the platform automatically registers it in the model registry, updates metadata, and triggers molecular database precomputation to re-score existing compound libraries—ensuring discovery organizations continuously benefit from accumulating experimental data without manual model maintenance.

*Client Integrations*

Client Integrations extend the platform's capabilities to encompass enterprise-scale batch precomputation, integration with molecular design environments, and agentic interfaces enabling conversational access to computational infrastructure.

*Molecular Database Precomputation*

Molecular database precomputation provides automated batch inference for enterprise molecular libraries, triggered by model version updates or molecular database expansion. Precomputed predictions are cached with compound identifiers, model versions, and timestamps, enabling instantaneous retrieval for virtual screening campaigns and compound prioritization workflows—dramatically reducing latency for high-throughput scenarios where established models are repeatedly applied to known compound collections.

*Molecular Design Suite Integrations*

Integration with molecular design suites enables computational scientists to access predictive models directly within design environments without context switching. The platform provides API endpoints connecting to commercial and artificial intelligence (AI)-native molecular design applications across diverse modalities. Scientists can generate novel compounds within design suites, invoke the Model Gateway API with Simplified Molecular Input Line Entry System (SMILES) strings or amino acid sequences, and receive inline predictions for absorption, distribution, metabolism, excretion, and toxicity (ADMET) characteristics and developability. The asynchronous architecture ensures design applications remain responsive even for computationally expensive models.

*Agentic LLM Integrations*

Agentic LLM integration enables scientists to interact with predictive infrastructure through natural language rather than programmatic APIs. This manuscript does not describe the design of LLM agents themselves; rather, this section illustrates how Model Gateway exposes a standards-compatible tool surface that any conforming agent can consume. The Model Gateway exposes its model catalog as a machine-readable tool surface compatible with agentic frameworks (e.g. vendor-native function calling and the Model Context Protocol, MCP [35]), so that an enterprise chat agent or a third-party chemistry agent[14,15,17,18] can list, introspect, and invoke models without bespoke wrappers per integration. The metadata-driven approach ensures agents automatically gain awareness of newly registered models without re-engineering.

To illustrate end-to-end tool-calling behavior, consider a multi-step natural-language request issued through an internal enterprise chat assistant:

> *"Find molecules similar to this SMILES, estimate their ADME properties, and filter out any with LogD greater than the reference compound."*

The agent decomposes this into three sequential tool calls against the Model Gateway: (1) a similarity-search invocation that returns a ranked list of structurally analogous compounds from the enterprise library; (2) a batch ADME submission for the full candidate set, where each molecule is submitted asynchronously and collected under a single workflow UUID, returning predictions such as those in Table 3; and (3) a filtering step in which the agent compares each compound's predicted LogD against the reference value and discards those above the threshold, returning a ranked, filtered table ready for experimental prioritization. Each step's model name, version, and workflow UUID are logged to S3 for reproducibility and audit. This pattern—tool decomposition, asynchronous batch execution, result-conditional filtering—is the practical utility of exposing a governed model catalog as an agent-native tool surface, and is not achievable with bespoke per-model wrappers.

As multi-tool workflows grow in complexity, a new reliability challenge emerges: agents can select semantically overlapping tools when the catalog is large and tool descriptions share vocabulary. This is not a gateway failure per se, but it motivates continued investment in metadata specificity and capability tagging, and is discussed further in Section a.

### Simulated Performance Example

To illustrate platform scalability, we present a simulated performance example using a sandbox deployment with 16 virtual central processing units (vCPUs), 64 GB memory, 16 client workers, and a consensus model with 1-second execution time per prediction. Production inference telemetry is omitted as compound library sizes and usage patterns constitute proprietary information; the sandbox deployment isolates platform behavior from model-specific variability and enables reproducible characterization of the queuing and scaling architecture. The deployment utilized Redis Cluster with three master and three worker nodes and PostgreSQL for model metadata storage. Load testing used the Locust framework, progressively scaling from 1 to 100,000 concurrent users, as illustrated in Fig. 3.

Job submission maintained 100% success up to 10,000 concurrent users; beyond this threshold, queue saturation caused success rates to decline under fixed sandbox resources (Fig. 3). Result retrieval achieved 100% success at all concurrency levels due to the lower overhead of cache reads versus job enqueuing. In production, KEDA-driven autoscaling shifts these saturation points dynamically by provisioning additional worker replicas as queue depth grows.

### Impact and Adoption

The Model Gateway is the production model-serving fabric for Lilly drug discovery and currently provides access to over 200 models with model inferences on the order of 100 Million per month. These models span a variety of modalities including small molecule, peptide, and antibody modalities, with oligonucleotide and conjugate modalities in active development. These models actively support the Design, Test, and Analyze phases of the DMTA cycle across all active small-molecule and large-molecule discovery programs, and are consumed by more than five downstream user-facing applications including molecular design suites, an internal enterprise chat assistant, the Model Gateway web user interface (UI), and automated batch precomputation pipelines that re-score the enterprise compound collection on every model update.

The most directly observable impact has been on *model traceability and governance*. Prior to Model Gateway, model weights accreted across shared HPC folders into more than 10,000 artifacts with substantial functional overlap and uncertain provenance. Under the platform, all models—including physics-based and cheminformatics tools previously outside any registry—are centrally registered, versioned, and reproducible. Because every inference request is keyed to a workflow UUID, model name, and explicit version, scientists can compare predictions across model releases deterministically, access controls are auditable at the individual model×user level, and the question "what model should I use, and which version?"—previously answered by tribal knowledge—has a governed, enforceable answer. Defining the heuristics by which project models graduate to global status remains an active collaboration between platform engineering and the discovery business.

In pharmaceutical drug discovery, predictive models are not the rate-limiting step in DMTA cycles—wet-lab synthesis and assay execution are. The value of a governed serving platform therefore lies not in accelerating cycle time directly, but in the *quality and reproducibility of the computational decisions that inform which experiments to run*. Model Gateway contributes to this through four mechanisms that are difficult to recover from ad hoc model deployments: (1) every prediction is keyed to an explicit model version and assay training snapshot, so scientists comparing predictions across design iterations are always comparing like-for-like; (2) global ADMET and property models are retrained on a monthly cadence against the most recently synthesized Lilly compounds, ensuring predictions remain calibrated to the expanding chemical space of active programs; and (3) by surfacing the model catalog at the application layer—molecular design suites, an enterprise chat assistant, and the platform web UI—the platform reduces the friction of model access for non-computational scientists who previously had no path to these tools. (4) by standardizing model declaration & investing in the scaled execution layer, we can leverage the same model to provide low latency access across design tools while also supporting large scale batch inferences. The marginal retrospective benefit relative to prior ad hoc patterns is inherently difficult to quantify; building a telemetry layer that can characterize model influence on experimental decisions is among the platform's highest-priority next developments.

### System Evolution and Lessons Learned

Several aspects of the Model Gateway have evolved since the first production deployment, and we summarize the most important learnings here in the spirit of an experience report.

Fig. 4 summarizes the platform's evolution from initial deployment through current production capabilities and planned enhancements.

*Real-world telemetry is hard.* Operationalizing defensible retraining heuristics across heterogeneous model types has proven nuanced and challenging; no single decay metric travels well across model families. The platform currently supports time-based retraining schedules, with model-by-model drift detection policies an active area of development.

*RBAC must follow the science, not the file system.* Our initial access model was loosely modeled on filesystem permissions and had to be redesigned around the actual organizational units of drug discovery—programs, therapeutic areas, and modality-specific computational groups—once we began onboarding cross-functional teams. The current three-tier scheme (Superuser, Model Owner, Reader) and the convention that models are private by default with explicit grants is a direct response to that experience.

*Asynchrony is non-negotiable.* Synchronous inference paths were convenient for the initial small set of millisecond-latency models but became a liability the moment physics-based and consensus models entered the registry. Re-architecting around asynchronous submission and retrieval leveraging Redis to hold hot state across model execution nodes was the highest-leverage change in the platform's history and is now a hard contract for new model onboarding.

*LLM agents need a metadata-driven catalog, not a hand-curated one.* Early agent integrations were maintained as bespoke wrappers per model, which scaled badly and fell out of sync with the registry as models were updated or replaced. Switching to a metadata-driven tool surface (and an MCP-compatible interface [35]) decoupled agent development from model onboarding entirely: any newly registered model becomes immediately reachable by every agent that consumes the catalog, with no wrapper maintenance required.

*Adoption requires meeting scientists at their existing workflows.* Centralized governance initially encountered resistance from computational scientists accustomed to direct model-file access on the corporate HPC: routing predictions through an API felt like trading control for overhead. The platform addressed this by providing a command-line interface that preserved familiar scripting patterns while adding versioning and audit logging, alongside application-layer integrations that gave non-computational scientists their first programmatic path to predictive models. Serving both constituencies from a single backend—with governance consistent across both access modes—proved more effective than asking either group to substantially change their working habits.

*Complex agentic workflows surface tool-selection reliability challenges.* As multi-tool workflows have grown in length and ambiguity—chaining similarity search, ADMET prediction, filtering, and consensus scoring from a single conversational request—a new failure mode has emerged: agents occasionally select semantically overlapping tools when the catalog is large and tool descriptions share vocabulary. This is not a gateway failure per se, but it has motivated ongoing investment in improved model metadata specificity, capability tagging, and agent-side tool-disambiguation guidance. Robust agentic reliability at scale remains an active and unsolved problem across the field[18,19], and the design of agent orchestration layers that gracefully handle large, heterogeneous tool catalogs is an important direction for future work.

## Conclusion

The Model Gateway addresses critical MLOps challenges in pharmaceutical drug discovery by providing a production-grade, cloud-based platform for end-to-end model lifecycle management. The platform's design directly addresses pharmaceutical-specific requirements: hierarchical RBAC enables organizational delineation across heterogeneous discovery teams; flexible version management accommodates both ML models through MLflow and physics-based computational methods through custom versioning; metadata-driven model discovery enables LLM agents—and emerging industry-standard tool surfaces such as MCP [35]—to autonomously select and execute appropriate models through natural language interfaces; asynchronous execution supports variable latency from milliseconds to hours; elastic scaling with heterogeneous GPU/CPU instance types enables dynamic resource adaptation; and consensus model management implements MPO strategies for compound prioritization across competing molecular properties. In production at Eli Lilly, the platform now governs more than 200 actively deployed models—a consolidation from over 10,000 unmanaged weight artifacts on shared HPC—and serves more than five downstream applications across all phases of DMTA. By operationalizing MLOps principles specifically for pharmaceutical workflows, the Model Gateway reduces technical debt, increases model accessibility, ensures reproducibility, and accelerates the translation of computational innovation into drug discovery impact. Planned enhancements include production model performance analytics and drift detection, predictive container warming, extended automated retraining for graph neural networks and transformer-based biologics models, and declarative multi-model pipeline orchestration.

## Acknowledgment

We thank all leadership and colleagues across the organization who supported, contributed, and provided resources for this project.

## CRediT Author Contribution Statement

**Yan-Shiun Wu:** Conceptualization, Methodology, Software, Writing – Original Draft, Writing – Review & Editing, Project Administration. **Sai Mahit Vaddadi:** Software, Methodology, Writing – Review & Editing. **Zachary A. Rollins:** Conceptualization, Writing – Original Draft, Writing – Review & Editing. **Nathan A. Morin:** Conceptualization, Supervision, Writing – Review & Editing, Project Administration.

### Declaration of Generative AI and AI-Assisted Technologies in the Writing Process

During the preparation of this work the authors used AI-assisted tools to support manuscript editing and formatting. After using these tools, the authors reviewed and edited the content as needed and take full responsibility for the content of the published article.

### Data Availability Statement

No datasets were generated or analysed during the current study. Simulation configurations and parameters used in the performance evaluation are available from the corresponding authors upon reasonable request.

Table 1: Capability comparison of representative platforms against the Model Gateway. ✓ = supported and described; **P** = partial or not a headline capability; − = not described or not applicable. Columns reflect capabilities that differentiate pharma enterprise MLOps from general-purpose or domain-specific alternatives. *Pharma RBAC* : access control structured around scientific org units (programs, therapeutic areas, modality groups). *Consensus/MPO*: inference-time composition of multiple models via weighted formula at the gateway layer. *Auto retrain*: automated retraining pipelines triggered by new experimental data. *Physics models*: registry and execution of non-ML artifacts (docking, molecular dynamics (MD), cheminformatics executables). *Compound precomp.*: batch re-scoring of an enterprise compound library on model update. *Multi-client API* : designed as an adaptable vendor-neutral backend capable of exposing custom APIs interfaces unique to competing design environments. *LLM/agent*: metadata-driven discovery and invocation of models by LLM agents. Open/Prop./Comm. = open-source / proprietary internal / commercial. REINVENT 4 is included as a representative generative design tool; it is not a serving platform and scores accordingly. Agentic chemistry systems (ChemCrow, Coscientist, et al.) are consumers of a platform such as Model Gateway rather than alternatives to it, and are discussed in the text rather than scored here.

| **System** | **Pharma RBAC** | **Consensus/ MPO** | **Auto retrain** | **Physics models** | **Compound precomp.** | **Multi-client API** | **LLM/ agent** | **Open/ Prop./Comm.** |
|---|---|---|---|---|---|---|---|---|
| *Pharma-internal platforms* | | | | | | | | |
| **Model Gateway (this work)** | ✓ | ✓ | ✓ | ✓ | ✓ | ✓ | ✓ | Prop. |
| AstraZeneca DMTA [6] | ✓ | P | P | P | P | – | – | Prop. |
| *MLOps suites* | | | | | | | | |
| Databricks Lakehouse LS [21] | ✓ | – | ✓ | – | – | ✓ | P | Comm. |
| Domino Data Lab [22] | ✓ | – | ✓ | – | – | ✓ | – | Comm. |
| NVIDIA BioNeMo [23] | – | – | P | P | – | P | – | Comm. |
| *Vendor scientific suites* | | | | | | | | |
| Schrödinger LiveDesign [24] | ✓ | ✓ | – | P | P | – | P | Comm. |
| Dotmatics Luma [25] | ✓ | – | – | – | – | – | – | Comm. |
| BIOVIA Pipeline Pilot [26] | ✓ | P | P | P | – | – | – | Comm. |
| Chemistry42 [20] | P | P | P | P | – | – | – | Comm. |
| *Open-source pharma tools* | | | | | | | | |
| NERDD [7] | P | – | – | – | – | P | – | Open |
| QSARtuna [8] | – | P | – | – | – | – | – | Open |
| REINVENT 4[13] | – | – | – | – | – | – | – | Open |
| ADMET-AI [9] | – | P | – | – | – | – | – | Open |

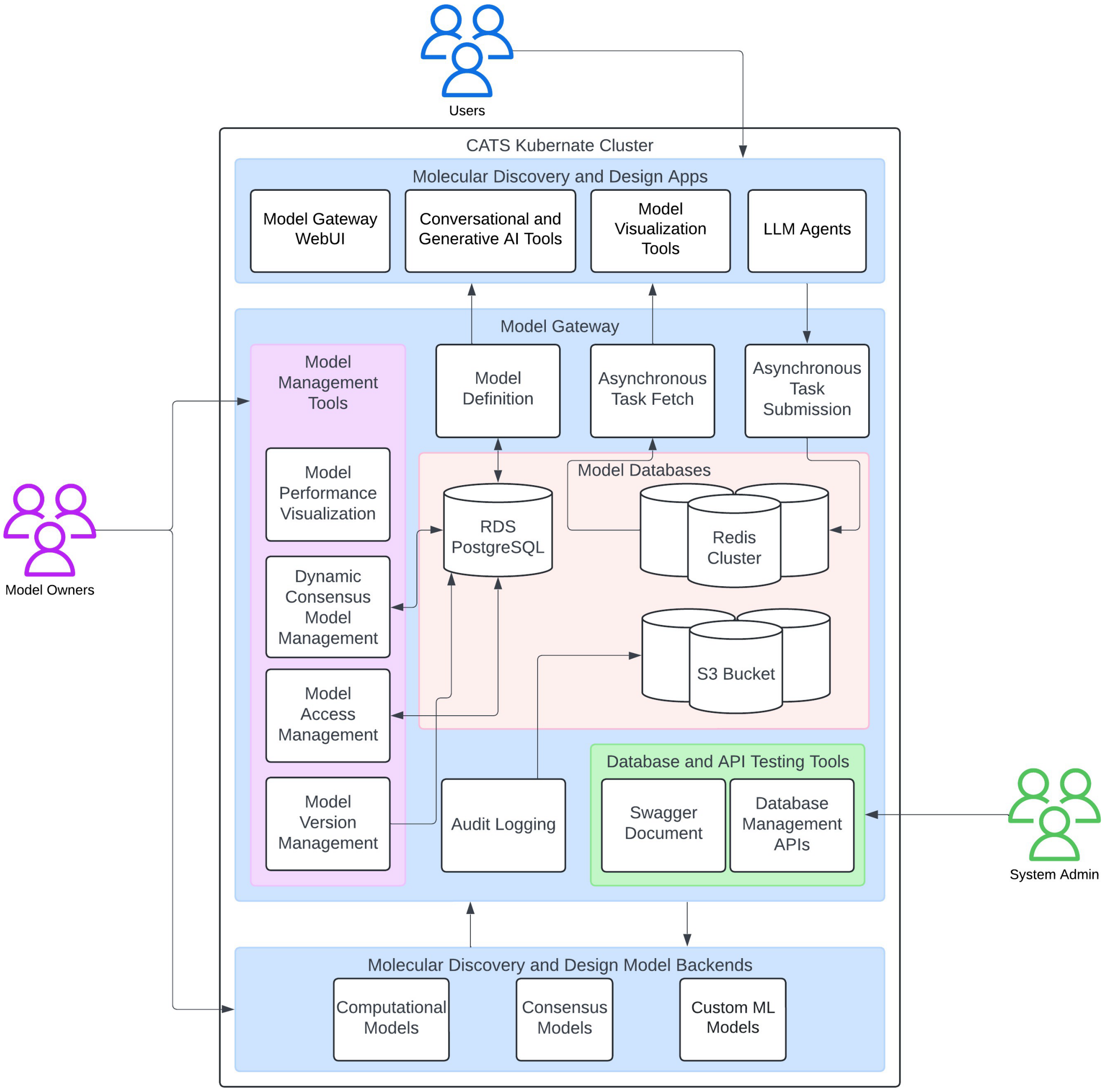

Figure 1: Model Gateway architecture. Client-facing applications (top) interact with the platform through a unified API within the Kubernetes cluster. The Model Gateway comprises a Model Management Tools panel for ownership, versioning, and consensus model configuration (left); a core execution layer for model definition and asynchronous job dispatch (centre); a Model Databases layer providing metadata storage (RDS), job queue (Redis), and artifact storage (S3) (upper right); and administrative testing and audit tools (lower right). Requests are routed to heterogeneous model backends (bottom).

Table 2: Component technology selections for the Model Gateway platform. Each row documents the requirement, options evaluated, the reason the alternative was not adopted as-is, and the decision taken.

| Requirement | Options evaluated | Why not adopted as-is | Decision |
|---|---|---|---|
| Model registry | MLflow [29], Weights & Biases [30] | W&B is Software-as-a-Service (SaaS)-hosted; both lack support for non-ML (physics-based) artifacts | Adopt MLflow; extend with custom version manager for physics-based models |
| Container orchestration | Kubernetes [27], Amazon Web Services (AWS) Elastic Container Service (ECS), on-prem high-performance computing (HPC) | ECS is AWS-only; HPC schedulers are poor fit for low-latency online inference | Adopt Kubernetes (Amazon Elastic Kubernetes Service, EKS [31]) for portability and ecosystem |
| Autoscaling | Central processing unit (CPU)-based Horizontal Pod Autoscaler (HPA), KEDA [32], manual provisioning | CPU HPA reacts only after pods saturate; manual provisioning wastes capacity for bursty workloads | Kubernetes Event-Driven Autoscaling (KEDA) queue-depth scaling (horizontal); heterogeneous GPU/CPU instance types per container (vertical) |
| Job queue | Redis [33], RabbitMQ, Amazon Simple Queue Service (SQS) | SQS is polling-based with higher latency and requires a separate store for result retrieval; RabbitMQ adds dedicated cluster management overhead without providing the key-value result caching needed for UUID-indexed job retrieval | Redis Cluster: queue, result cache, and UUID indexing in one component |
| Workflow orchestration | Airflow, Argo, Nextflow [34] | Airflow is Python-DAG-centric with no native per-step containerization; Argo requires verbose Kubernetes YAML and lacks a scientific dataflow programming model | Nextflow for retraining pipelines: channel-based dataflow model maps naturally to data-driven scientific workflows and has broad cheminformatics community adoption |
| LLM / agent access | Proprietary AI assistants (e.g. Microsoft Copilot), bespoke per-model wrappers, Model Context Protocol (MCP) [35] server | Proprietary assistants require bespoke per-vendor plugin integration, creating maintenance overhead and vendor lock-in; bespoke wrappers do not scale across a large model catalog | Metadata-driven tool surface compatible with any conforming agent (internal LLM, MCP client, or enterprise assistant) without per-integration maintenance |

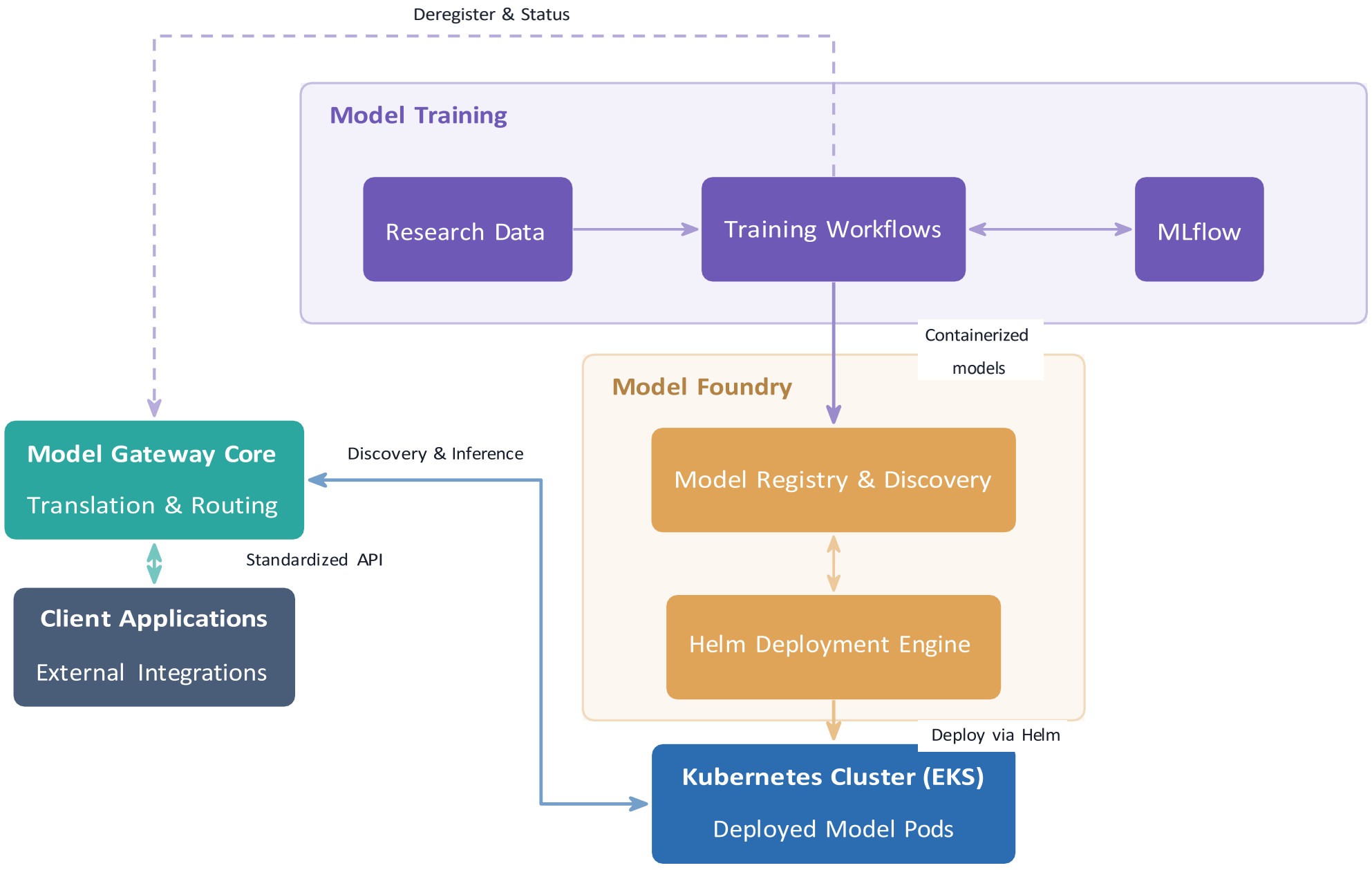

Figure 2: Model Gateway training and deployment architecture. Research data flows through training workflows tracked in MLflow (Model Training). Trained models are containerized, registered in the Model Foundry, and deployed via Helm to the Kubernetes cluster. The Model Gateway Core routes inference requests from client applications to deployed model pods, and receives deregistration and status signals from the training pipeline (dashed).

Table 3: Example ADME properties returned by the Model Gateway for a batch SMILES query (step 2 of the worked example). Each row is produced by an independently versioned model registered against the `adme` capability tag.

| **Property (model)** | **Prediction** |
|---|---|
| LogD (pH 7.4) | 0.38 |
| Madin-Darby Canine Kidney Permeability, Human (MDCKR) | 4.72 |
| Solubility at pH 7.4 (High-Throughput Solubility Assay, HTSA class) | >0.1 mg/mL |
| Human Hepatocyte $Cl_{int,u}$ (mL/min/kg) | 6.29 |
| Human Microsomal $Cl_{int,u}$ | 4.77 |

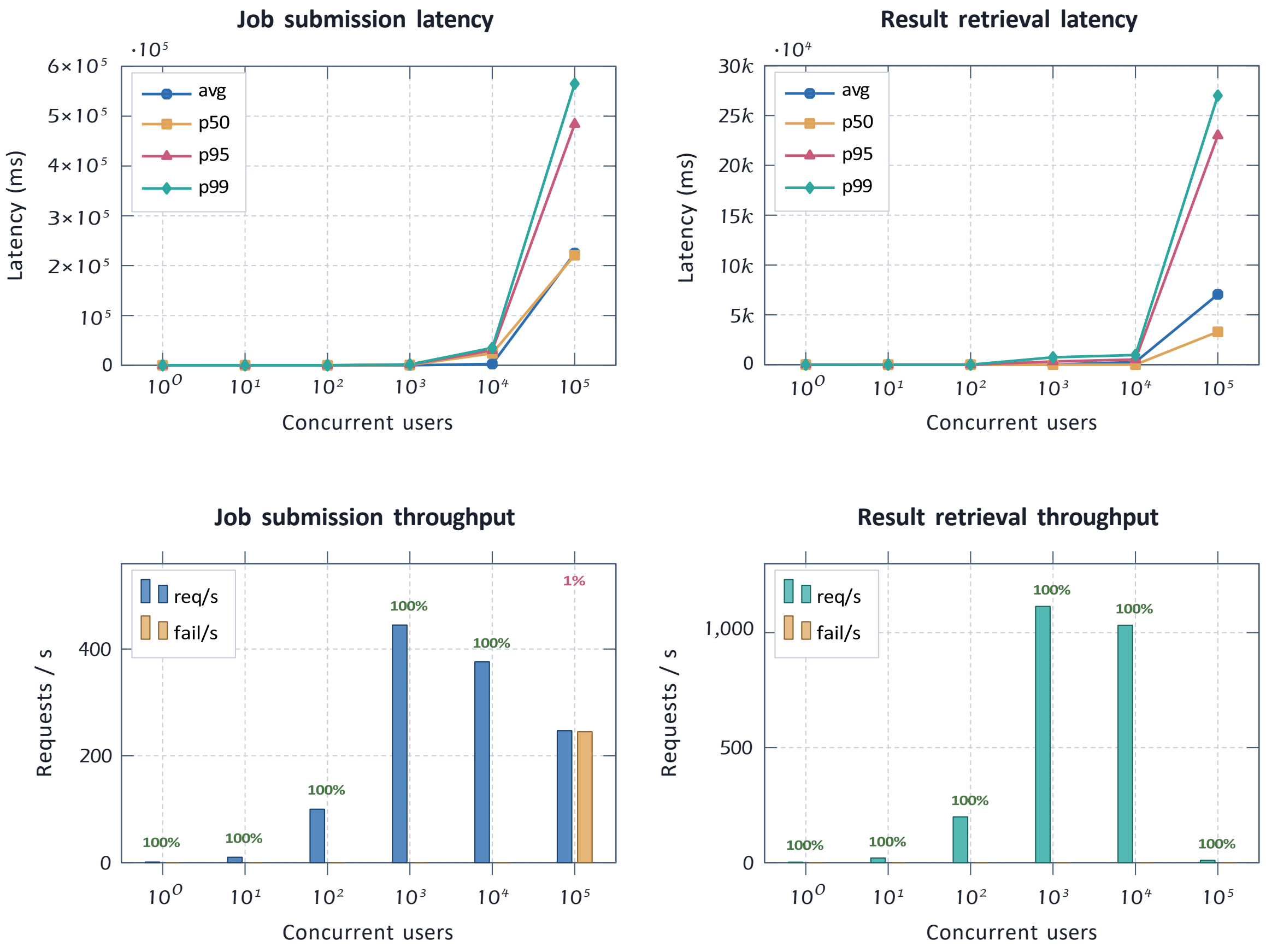

Figure 3: Simulated load-test profile across six concurrency levels ($10^0$ – $10^5$ users) on a fixed 16-vCPU sandbox. *Top row:* per-request latency percentiles (avg, p50, p95, p99) for job submission (left) and asynchronous result retrieval (right). *Bottom row:* sustained throughput (req/s) and failure rate (fail/s), with end-to-end success rate annotated above each load level. Submission saturates near $10^4$ users as the queue fills, falling to ~1% success at $10^5$; retrieval remains at 100% across all levels because cache reads bypass enqueue overhead. In production, KEDA-driven autoscaling shifts these saturation points by provisioning additional worker replicas as queue depth grows.

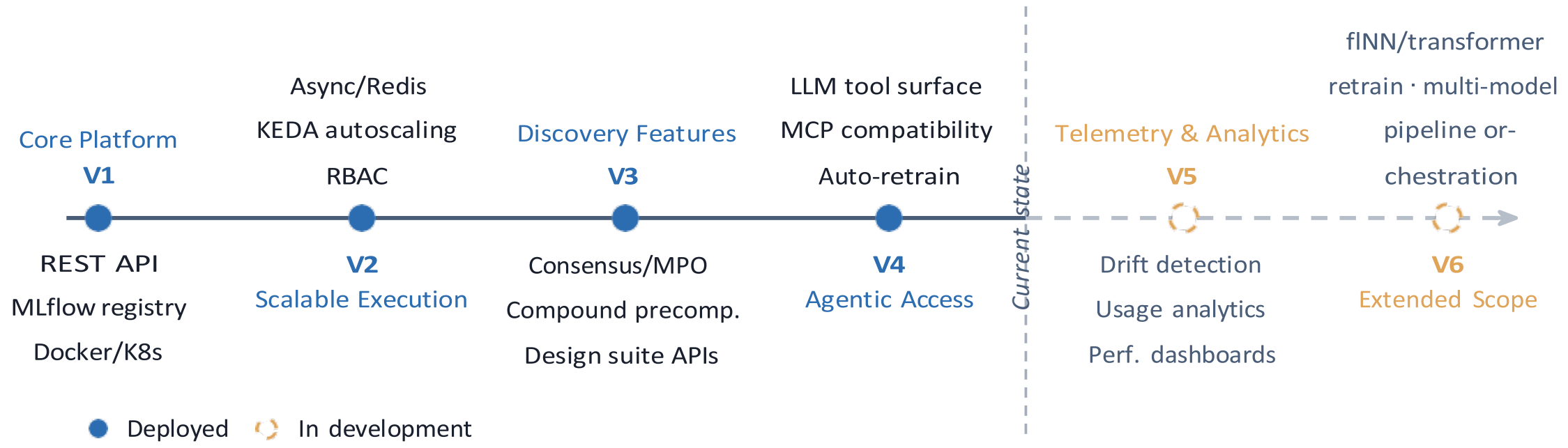

Figure 4: Model Gateway evolution timeline. Solid blue nodes represent capabilities in production; open amber nodes represent planned enhancements. The dashed vertical line marks the platform state at time of writing.